# Plasmonic Dimers Enhanced Polarized Single Photon Coupled to Optical Nanowire

Subrat Sahu[1], Harsh Mishra[2], and Rajan Jha[1*]

[1] Nanophotonics and Plasmonics Laboratory, School of Basic Sciences, IIT Bhubaneswar, Khurda, 752050, Odisha, India
[2] Department of Physical Sciences, Indian Institute of Science Education and Research, Berhampur, 760010, Odisha, India

(*Electronic mail: rjha@iitbbs.ac.in)

We propose a system for guiding plasmon-enhanced polarized single photons into optical nanowire (ONW) guided modes. It is shown that spontaneous emission properties of quantum emitters (QEs) can be strongly enhanced in the presence of gold nanorod dimer (GNRD) leading to the emission of highly polarized and bright single photons. We have calculated that a high Purcell factor of 279, coupling efficiency of 11 %, and degree of polarization (DOP) of single photons is estimated to be as high as 99.57% in the guided modes of ONW by suitably placing a QE on an optimized location of the GNRD system. This proposed hybrid quantum system can be in-line with fiber networks, opening the door for possible quantum information processing and quantum cryptography applications.

Generation and detection of polarized single photons (PSP) at room temperature have been in high demand, to carry out quantum information processes[1]. So efficient generation, channeling, and manipulation of PSP are of utmost importance in contemporary quantum optics[2]. Quantum emitters (QE) like solid-state quantum dots (QD)[3], atom[4], and defect center[5,6] have been reported for generating single photons. But these QEs are showing emission bleaching and blinking effect, low photon count rate, and less polarized[7]. For an efficient QE, single photons generated from it should have a high photon count rate and a high degree of polarization (DOP), because PSPs are used for quantum cryptography applications[8]. So, an effective way to generate bright and PSP from a QE, at room temperature is through the localized surface plasmon resonances (LSPRs) using plasmon active metal nanostructures like nanorods (NR) and nanoparticles (NP)[9]. Because of their intriguing optical characteristics, the conduction electrons in metal nanostructures coherently oscillate with the incident light, creating an enhanced electric field that is localized over the surface of the nanostructure and makes it feasible to confine light in all three dimensions at a deep-subwavelength scale. Hence, in this prospect, a metal NR is the best candidate as compared to a metal NP, because the aspect ratio of NR can be changed, so the operating wavelength can be tuned as per the requirement[10].

On the other hand, a key aspect is to efficiently channel the PSP to the nanostructure waveguides in-line with existing optical fiber networks (single mode fiber, SMF) for different applications. The necessary technology to develop compact quantum systems is being realized using nanostructured optical waveguides and resonators that offer efficient advantages over free space-based systems[11]. So, a subwavelength diameter SMF called an optical nanowire (ONW), is one of the most suitable platforms to the couple and guide such PSP. In this type of cylindrical dielectric waveguides, light exits outside of the fiber as an evanescent field, which makes it easier to interface with the emitter[12], and also with the metal nanostructure[2]. It has been analytically estimated that 28% of the light can be channeled into the ONW-guided modes from a Cs-atom[13]. This ONW has excellent propagation properties for long-distance secured communication.

In this work, we propose a system for guiding plasmon-enhanced PSP into ONW-guided modes. It is shown that spontaneous emission properties of QEs can be strongly enhanced in the presence of gold nanorod dimer (GNRD) leading to the emission of highly polarized and bright single photons that can be efficiently coupled to guided modes of the ONW. We have calculated that a high Purcell factor of 279, coupling efficiency of 11 %, and DOP of single photons are estimated to be as high as 99.57% in the guided modes of ONW by suitably placing a QE on an optimized location of the GNRD system.

Figure 1 shows the schematic diagram of the hybrid quantum system formed by combining a single QE and GNRD over an ONW. The GNRD is formed by two equivalent gold nanorods (GNRs) of pure gold rods with hemispherical end caps placed very close to each other. The enlarged portion shows the waist region, having a diameter of ONW ($D_{ONW}$), each of the GNRs has length $L$, $h$ as diameter, and the distance between the two GNRs is $d$. With proper excitation of QE, single photons are emitted in all possible directions. A part of single photons is coupled into the guided modes of the ONW. But these QEs suffered from photobleaching and blinking, which leads to a low photon count rate. Hence, the spontaneous emission of the QE is largely enhanced when QE is placed between the GNRD due to plasmonic characteristics, called as Purcell effect[14,15]. Because the enhancement spectrum of the LSPR corresponding to the GNRD overlaps with the emission spectrum of QE, hence the single photons coupled in the guided mode of ONW from this hybrid system are highly polarized.

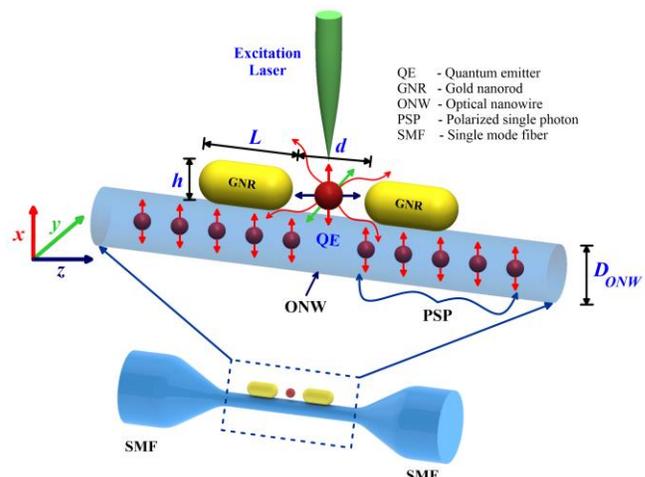

FIG. 1. Schematic illustration of the hybrid device having a QE+GNRD coupled system on an ONW. Enlarge portion shows the hybrid system near the waist region with different geometric parameters; $D_{ONW}$ = 300 nm, $L$ =122 nm, $h$ = 50 nm, and $d$ is the separation between the edge of two GNRs.

The complete analysis is done using finite-difference time-domain (FDTD) simulations. In this study, a system is designed assuming a QE-like colloidal quantum dot[16] or germanium defect center in nanodimond[17] having an emission wavelength of around 600 nm, due to the availability of an economical source. In this simulation, ONW is considered a cylindrical silica core with a refractive index of 1.450, having a diameter of $D_{ONW}$ with air cladding. The aspect ratio of GNR can be taken as, $AR = L/h$[18]. The dielectric constant of GNR is chosen in accordance with the fitting of experimental data from Johnson and Christy[19]. Here, an electric dipole source is chosen as a QE, emitting over a range of wavelengths from 500 to 700 nm, having three different polarizations (pol.), i.e., $x$-, $y$-, and $z$-pol. The analysis is performed for all three polarizations by changing the required angles of QE.

To study the absorbance of GNRD, the system is excited by a $z$-pol plane wave source wavelength ranging from 500-700 nm. Figure 2(a) shows the absorption spectra (dashed blue line) of GNRD, having a plasmon resonance peak around 595 nm with a full width at half-maximum (FWHM) of 58 nm[20]. Now, the photoluminescent (PL) spectra (solid red line) is recorded by placing a QE in the GNRD system as shown in Fig. 2(a). Here, PL spectra peaks around 601 nm with an FWHM of 33 nm. In order to understand the electric field enhancement due to the LSPR, the scattering field profile of GNRD at peak absorbance wavelength i.e., 595 nm is shown in the inset (up) of Fig 2(a). One can see that the field enhancement is maximum at the center of the GNRD forming a hotspot, so, this forms a strong plasmonic nanocavity between two GNRs. The electric field intensity ($E$) decreases as $d$ increases. The normalized electric field spectra of the scattered field is shown in the inset (down) of Fig. 2(a). As can be seen, there is a stronger field enhancement in the hotspot region. Now, if a QE emitting around 600 nm is placed at the maximum field region between the GNRD, then the spontaneous emission of the QE can be greatly enhanced[21].

Figure 2(b) shows the change in LSPR peak resonance wavelength ($\lambda_r$) as a function of $AR$ from 2.1 to 3.0. The solid lines represent linear fit to the points with a regression coefficient of $R^2 = 0.99$. Here for $h = 50$ nm, $L$ is varied from 105-150 nm and the LSPR spectra are recorded as shown in the inset of Fig. 2(b). The LSPR spectra show redshifts in $\lambda_r$ with increasing $AR$ of GNRs[10]. However, the effect of the LSPR mode is much more pronounced showing a variation of $\lambda_r$ over 105 nm as $AR$ changes from 2.1 to 3.0. By optimizing this $AR$, the dimension of each GNR is chosen as $L = 122$ nm, $h = 50$ nm, and $AR = 2.44$ as the PL intensity spectra peak around 600 nm at this dimension.

Further, to see the enhancement properties of GNRD, the fluorescence enhancement factor (FEF) of GNRD is calculated as the ratio of the average electric field intensity between the longitudinal ($z$-pol) and transverse ($y$-pol) polarization excitations of plane wave source with the GNRD axis[22]. Figure 2(c) shows the calculated average electric field intensity enhancement (pink circles) in the GNRD plotted as a function of the excitation polarization angle ($\theta$) from $0 - 2\pi$. As can be seen that there maximum (minimum) field enhancement observed at $\theta = 0$ ($\pi/2$) corresponds to longitudinal (transverse) polarization excitation. The field intensity enhancement can be well-fitted (green line) with, $E/E_0 = A\cos^2\theta + B$[22], where A = 45 and B = 0.86. The small enhancement value of 0.86 is calculated under the transversely polarized excitation. Moreover, the FEF is calculated as 52.3.

Further, to see the spontaneous emission characteristics of QE in presence of GNRD; Purcell factor ($P_f$) with the emission enhancement factor (EF) of QE, and coupling efficiency ($\eta$) with DOP ($P$) of the emitted photons into the ONW guided modes are calculated by placing a QE in between two GNRs of the GNRD configuration on an ONW. Here, the QE has placed 4 nm away from the surface of ONW by considering the experimental realization of the size of a solid-state quantum emitter like QDs, defect centers, and 2D materials[3].

Now, the total spontaneous emission decay rate ($\gamma^T$) of a QE near a hybrid system can be written as $\gamma^T = \gamma^g + \gamma^r$, where $\gamma^g$ is the emission rate of QE into the guided mode of ONW and $\gamma^r$ is the emission rate of QE in the radiation mode. Here, the $P_f$ is determined by $\gamma^T/\gamma^0$, where $\gamma^T$ ($\gamma^0$) is the total decay rate in the presence (absence) of ONW and GNRD. The EF of QE is calculated as $\gamma^c/\gamma^{un}$, where $\gamma^c$ ($\gamma^{un}$) is the spontaneous emission of the emitted light coupled to the guided modes of ONW in the presence (absence) of the GNRD, respectively. The $\eta$ for both sides of the guided modes are determined as $\gamma^c/\gamma^T$. And, the DOP is determined by $P = (I_x + I_z - I_y)/(I_x + I_z + I_y)$, where $I_m$ ($m = x, y, z$) is the proportion of the emitted light intensity coupled to the ONW for a QE polarized along the $m$-axis[2,21].

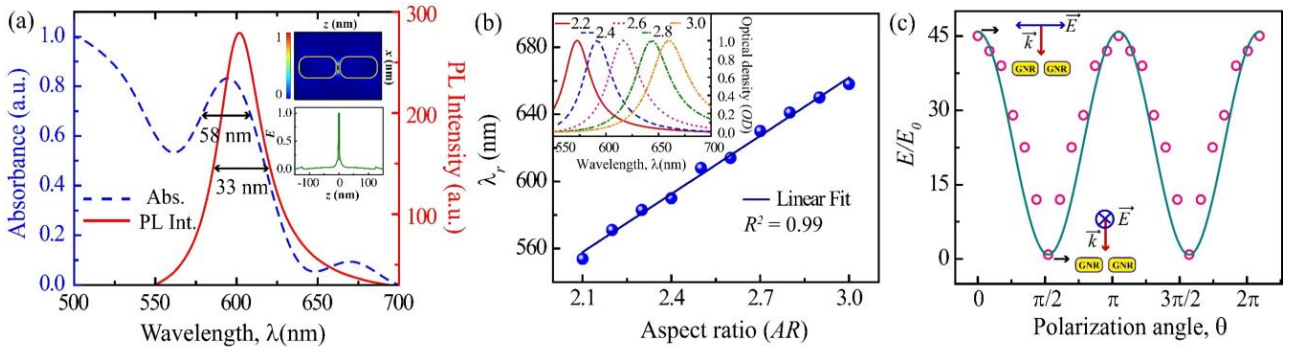

FIG. 2. (a) Shows the normalized absorbance spectra (dotted line) of GNRD and PL intensity spectra (solid line) of the GNRD+QE. The insets show the scattering field profile (up) and field spectra (down) of GNRD. (b) Change in resonance wavelength ($\lambda_r$) for different aspect ratios (AR) of GNRD. The inset shows the variation of LSPR spectra of GNRD with different AR. (c) Field intensity enhancement (pink circles) as a function of the excitation polarization angle ($\theta$). The green line is a fit.

Now, one needs to optimize $D_{ONW}$ to maximize the η of plasmon-enhanced highly PSP into the guided modes of ONW. Figure 3(a) shows the variation in $P_f$ and η with the $D_{ONW}$ for three different orthogonally polarized QE, i.e., $x$-, $y$-, and $z$-pol QEs. One can readily see that the small oscillations observed in $P_f$ by varying $D_{ONW}$, this is due to the properties of Green's dyadic function[23] and these are called Drexhage oscillations[24]. On the other hand, η initially follows an increasing trend and then a decreasing trend with a change in $D_{ONW}$ in all polarized QEs, this is due to the appearance of additional guided modes as $D_{ONW}$ increases. Now, the maximal η of 11%, 0.66%, and 1.3% are realized for $x$-, $y$-, and $z$-pol QE, respectively at a $D_{ONW}$ of 300 nm, a value of almost half the emission wavelength ($\lambda$ = 600 nm). Figure 3(b) shows the variation of $EF$ and $P$ as a function of $D_{ONW}$. One can see that maximum $EF \sim 31$ and $P = 99.57$ % are achieved around $D_{ONW}$ = 300 nm. Both of these responses mimic η-value.

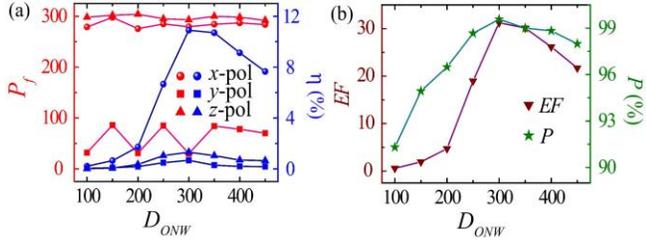

FIG. 3. (a) Shows the change in $P_f$ and η with the $D_{ONW}$ for $x$- (sphere), $y$- (square), and $z$-(triangle) pol QE, respectively. (b) Variation of $EF$ (triangle) and $P$ (star) for single photons propagating in the guided mode of the ONW as a function of the $D_{ONW}$.

Now to see the effect of "$d$" on the spontaneous emission characteristics of coupled QE, $P_f$ and η are plotted as a function of the distance $d = 3 - 103$ nm, between the edges of two GNRs, as shown in Fig. 4(a). One can readily see that $P_f$ is maximum at $d = 3$ nm with values around 279, 30, and 292 for $x$-, $y$-, and $z$-pol QEs, respectively. And then the $P_f$ rapidly decreases as $d$ increases. The $P_f$ for $z$-pol QE is maximum because spontaneous emission with polarization lying along the GNRD length axis ($z$-axis) will be principally enhanced[2,21]. It is found that quenching behavior is observed if the QE is in contact with GNRD[15]. On the other hand, η shows an increasing trend as $d$ increases for all QEs as shown in Fig. 4(a). Maximum $\eta \cong 11\%$ (20.4%) is achieved at $d = 3$ nm (103 nm) for $x$-pol QE because when the QE is oriented radially, maximum emitted photons from the QE couple into the guided modes of ONW[25]. Furthermore, η for $z$-pol QE is less, because the $z$-pol QE couples the most weakly to the ONW guided modes, out of the three QEs [25].

Figure 4(b) shows the variation of the $EF$ and $P$ for PSP propagating in the guided modes of the ONW as a function of $d$. As can be seen, the $EF$ shows a decreasing trend with an increase in $d$, and it becomes minimum at $d = 103$ nm. The $P$ also shows a decreasing trend with a value in the range of 99.57 – 75 % as $d$ increases. Apart from the high $P_f$, there is also a high value of $P$ observed in the hybrid system, but η is not in general enhanced[2,21] for a coupled QE system. The $P_f$, η ($P$) for uncoupled QE is also calculated as 1.256, 30 % (33.35 %) into the guided modes of ONW. As can be observed that $P_f$ and $P$ are very much less as compared with the coupled QE, though η is more in the case of uncoupled QE. So that if a QE is placed very close to GNRD, high PSPs are emitted having reduced decay time, enabling excitation with a high repetition rate, suppressing blinking effect leading to brightness, and can be operated in room temperature conditions [2,9].

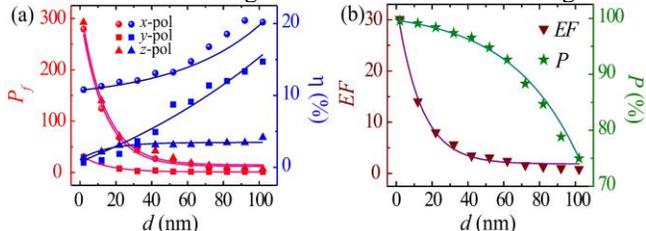

FIG. 4. (a) Shows the change in $P_f$ and η with $d$ between two GNRs for $x$-, $y$-, and $z$-pol QEs. (b) Variation of the $EF$ and $P$ for photons propagating in the guided mode of the ONW as a function of $d$.

Further, the dipole angle orientation of the QE is also affecting the spontaneous emission properties. The significant changes in polarization dependence of QE can be observed in Fig. 5(a), which shows the variation of $P_f$ and η with QE angle ϕ. One can see that both the parameters follow a decreasing trend with decreasing in the ϕ. Here, the QE angle ϕ = 90° (0°) corresponds to $x$-pol ($y$-pol) QE, respectively. From this graph, spontaneous emission characteristics of a randomly polarized QE can be determined. Here, maximum $P_f$ (η) can be obtained as 279 (11%) and 30 (0.66%) for $x$- and $y$-pol QEs, respectively. Figures 5(b) and 5(c) show the contour plots of the $P_f$ and $I^c$ over a range of wavelength 550-700 nm for randomly polarized QE. So, one can readily see that in both cases field distributions are maximum at ϕ = 90° and minimum at ϕ = 0°. Also, the broadening in the field distribution is observed as ϕ changes from 90° to 0°.

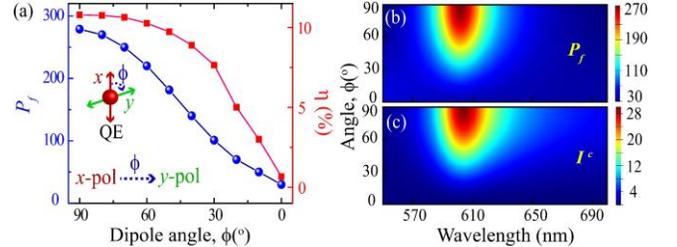

FIG. 5. (a) Change in $P_f$ and η, as the dipole polarization angle (ϕ) changes from $x$-pol to $y$-pol. (b) and (c) Show the contour plots for $P_f$ and $I^c$ by varying the ϕ.

Now, for experimental realization, it is very challenging to precisely place a QE between two GNRs on an ONW. But, many well-established techniques are developed to interface a single QE and GNR on an ONW[2,21]. In this hybrid system orientation of GNRs along the ONW axis is another challenge. But it is found that GNRs are always aligned almost parallel to the ONW axis due to the surface forces[26]. Further, accurate and deterministic positioning of QE and GNRs on the interface of ONW can be done by using a sophisticated AFM technique[27,28].

**Table 1: Performance comparison of reported works.**

| Ref. | $P_f$ | η (%) | DOP, $P$ (%) |
|---|---|---|---|
| Sugawara et.al. [21] | 88 | 6.8 | 86 |
| Shafi et.al. [2] | 60 | 3.2 | 94-97 |
| This work | 279 | 11 | 99.57 |

The performances of different types of works are summarized in table 1. From the comparison, one can infer the followings.

- By placing the QE near the edge of GNR as in[21], high Purcell enhancement occurs due to the LSPR effect, and η also increases due to the QE placed on the ONW.
- When QE is placed over the GNR as in [2], $P$ increases but the η decreases, because in this case, QE is not directly in contact with the ONW. So, it is suspected that maximum photons go in radiation mode.
- But, in our proposed system the QE is placed in a high electric field enhancement hotspot region of GNRD also it is in contact with the ONW. So, high Purcell enhancement occurs as well as η also increases.
- From the comparison with other works [2,21], we get higher $P_f$ and much better η with higher $P$ of PSP in the guided modes of ONW, in our proposed system.

Hence, our present study can be used for potential applications in the field of quantum information processing[1], because such a high $P_f$ corresponds to a high photon count rate which leads to high data rates. Also, extremely high $P$ of PSPs can be used in quantum cryptographic applications, for secured long-distance communications of encoded information. Now, such a high $P_f$ is achieved in this type of coupled system due to the plasmon enhancement of GNRs, which is in contrast to the cavity-based systems[29]. The η of plasmon-enhanced highly PSP in the guided modes can be further improved by making gratings on the ONW[30,31].

In conclusion, we propose a system for guiding plasmon-enhanced PSP into ONW-guided modes. It is shown that spontaneous emission properties of QEs can be strongly enhanced in the presence of GNRD leading to high PSP. We have calculated that a high Purcell factor of 279, coupling efficiency of 11 %, and DOP of single photons are estimated to be as high as 99.57% in the guided modes of ONW by placing a QE near the GNRD system. These PSPs can be transferred over a long distance in the conventional SMF network. Apart from applications for quantum communications, this system paves the way for a promising platform for the quantum plasmonic field.

**Acknowledgments:** RJ acknowledges the support of the SERB STAR Fellowship (STR/2020/000069) from Govt. of India.